\def\year{2016}\relax
\documentclass[letterpaper]{article}
\usepackage{aaai16}
\usepackage{times}
\usepackage{helvet}
\usepackage{courier}
\usepackage[capposition=top]{floatrow}
\usepackage{multirow}
\usepackage{caption}
\usepackage{amsmath}
\usepackage{graphicx}

\title{\Large Deciphering the 2016 U.S. Presidential Campaign in the Twitter Sphere: A Comparison of the Trumpists and Clintonists}
 
\begin{document}

\author{Yu Wang\\Political Science\\University of Rochester\\Rochester, NY 14627\\ywang@ur.rochester.edu \And Yuncheng Li\\Computer Science\\University of Rochester\\Rochester, NY 14627\\yli@cs.rochester.edu \And Jiebo Luo\\Computer Science\\University of Rochester\\Rochester, NY 14627\\jluo@cs.rochester.edu}
\nocopyright{}
\maketitle

\begin{abstract} 

In this paper, we study follower demographics of Donald Trump and Hillary Clinton, the two leading candidates in the 2016 U.S. presidential race. We build a unique dataset \textit{US2016}, which includes the number of followers for each candidate from September 17, 2015 to December 22, 2015. \textit{US2016} also includes the geographical location of these followers, the number of their own followers and, very importantly, the profile image of each follower. We use individuals' number of followers and profile images to analyze four dimensions of follower demographics: social status, gender, race and age. Our study shows that in terms of social influence, the Trumpists are more polarized than the Clintonists: they tend to have either a lot of influence or little influence. We also find that compared with the Clintonists, the Trumpists are more likely to be either very young or very old. Our study finds no gender affinity effect for Clinton in the Twitter sphere, but we do find that the Clintonists are more racially diverse.

\end{abstract}

\section{Introduction} 
Barack Obama's masterful use of social media in the 2008 presidential campaign has been widely credited as having established Twitter and other social media sites as integral parts of the political campaign tool box \cite{tumasjan}. Using Twitter, candidates can reach out to more voters and receive immediate feedback through, for example, the `likes' mechanism \cite{trumponfire}. They can also use Twitter to attack other candidates, which also helps them win votes. For example, Donald Trump attacked fellow Republican candidate Jeb Bush with the tweet:``@qbeacademy: Well I'm at 42\% and your at 3\% .You know Jeb you started over here you keep on moving further and further off stage.  Classic."\footnote{The quoted tweet can be found at: https://twitter.com/\\realDonaldTrump/status/679416309545873409.}

\begin{figure}[h!]
\caption{Clintonists in the Twitter Sphere\footnote{Photos retrieved from Twitter.}}
\label{gallery}
\includegraphics[width=8.4cm]{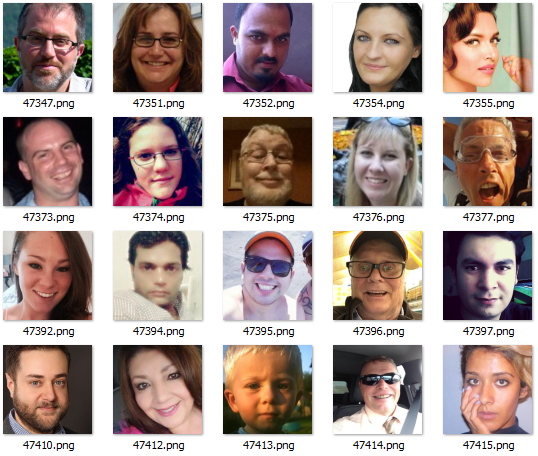}
\end{figure}

The focus of our work is on the Twitter followers of the two leading candidates: Donald Trump and Hillary Clinton. We refer to Clinton's followers as Clintonists and Trump's followers as  Trumpists (Figure \ref{gallery}). Our goal is analyze and compare the demographics of the two camps. First, tapping in the large sample size of \textit{US2016}, we study the social influence of Trumpists and Clintonists. We use the number of followers as a proxy for social status in the Twitter sphere. Second, by implementing a convolutional neural network (CNN) model on followers' profile images, we extract information on gender, race and age. We then study how the two camps differ in these dimensions.

Our study suggests that in terms of social influence the Trumpists are more polarized than the Clintonists: they tend to have either a lot of influence or little influence. Compared with the Clintonists, the Trumpists are also more likely to be either very young or very old. Our study finds no gender affinity effect for Clinton in the Twitter sphere, but we do find that the Clintonists are more racially diverse.

Our work contributes to the study of campaign effectiveness and to the study of Donald Trump's rise in American politics.
\section{Literature Review}
Our work builds upon previous research in electoral studies and in computer vision.

In eletoral studies, researchers have found that race and gender constitute two important factors in voting behavior. One common observation is that women tend to vote for women, which is usually referred to as gender affinity effect \cite{sexAndGOP,genderAffinityEffect}. In our work, we will test if women tend to follow Hillary Clinton on Twitter. As for race, analysis by \cite{amazingRace} shows that Barack Obama won the 2008 presidential election because of his support among African Americans. Using 2008 American National Election Studies survey data, \cite{obamaWhiteRace} shows that negative stereotypes about blacks significantly eroded white support for Obama. In this study, we will analyze the racial composition of the Clintonists and the Trumpists.




Our work also ties in with current computer vision research, as the profile images of the followers constitute an integral part of our \textit{US2016} dataset. In this dimension, our work is related to gender, race and age classification using facial features. \cite{facerace} provides a comprehensive survey of race classification based on facial features. \cite{israel} uses a five-layer network to classify both age and gender. Here we are inspired by \cite{ginosar} to explore the riches of followers' profile images. To our knowledge, our work is the first to use Twitter users' profile images for gender classification.


\section{Data Collection and Pre-Processing}
In this section, we describe our dataset \textit{US2016} and the pre-processing procedures. One variable is \textit{number of followers.} This variable is available for both candidates and covers the entire period from Sept. 17, 2015 to Dec. 22, 2015. For each candidate's followers we also have data on user name, number of followers, geographical information and profile images. 

In particular, we use the number of followers as a proxy for social status, assuming that individuals with a larger number of followers possess a higher social status.\footnote{A PageRank approach would require computing weights for each Twitter user. Here we adopt the simpler counting method.} We derive from the profile images the followers' demographic information, such as age, gender and race.

To process the profile images, we first use OpenCV to identify faces, as the majority of profile images only contain a face. We discard images that do not contain a face and the ones in which OpenCV is not able to detect a face. When multiple faces are available, we choose the largest one. Out of all facial images thus obtained, we select only the large ones. Here we set the threshold to 25kb. Lastly we resize those images to (256, 256). In Table \ref{image}, we report the summary statistics of the images in \textit{US2016}.

\begin{table}[]
\centering
\caption{Profile Images in 	\textit{US2016}}
\setlength{\tabcolsep}{7pt}
\label{image}
\begin{tabular}{llll}
\hline\hline
\multicolumn{2}{c}{Democratic Candidates} & \multicolumn{2}{l}{Republican Candidates} \\
\hline
Hillary Clinton         & 28, 771         & Ben Carson             & 20, 000          \\
Bernie Sanders          & 20, 000         & Donald Trump           & 29, 520        \\
\hline 
\end{tabular}
\end{table}

Our training data comes from MORPH database.\footnote{For summary statistics, please see http://people.uncw.edu/vetterr/MORPH-NonCommercial-Stats.pdf.} It has detailed labels on gender, race and age for 55,134 samples. This dataset will be used for training our CNN model. In the MORPH database, there are only 154 Asian faces. We decide to drop that class. In the training and validation stages, we only use White, Black and Hispanic images. So the classifier will only have these three classes.

For classifying images based on gender and race, we use the Caffe implementation of the ImageNet architecture that was pre-trained on the ILSVRC dataset \cite{imagenet}. Using the MORPH dataset, we find that the CNN model performs better than if we use either non-negative matrix factorization or principal component analysis to extract features and use SVM for classification. The Mean Average Precision (MAP) we achieve for gender classification is 99.8\%. For race, it is 95.6\%.\footnote{We randomly sample 40,000 images for training and use the rest for validation.} 

For lack of data to train the multiple age classes, we decide that for the purpose of age classification, we use the software service from Face++.\footnote{http://www.faceplusplus.com/.}
\section{Empirical Results} 
In this section, we first study the distribution of social influence of the two camps. We then investigate the gender, racial and age composition of Clintonists and Trumpists, and analyze their differences.

\subsection{Followers' Social Status}

In this subsection, we study the social influence of the Clintonists and the Trumpists. One way to measure social status is income, and there already exists research that predicts income based on user-generated tweets \cite{income}. Here we propose an alternative measure: the number of followers on Twitter. This is partly because we see a positive correlation between social status and the number of followers and partly because the subject of interest here is a follower of either presidential candidate.
\begin{figure}[H]
\caption{Trump leads in both Tails}
\includegraphics[height=4cm]{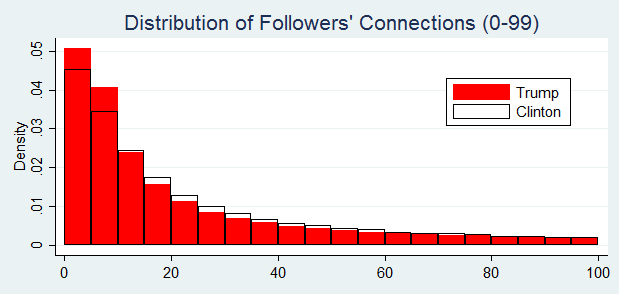}
\end{figure}

\begin{figure}[H]
\caption{Clinton leads in the Middle}
\includegraphics[height=4cm]{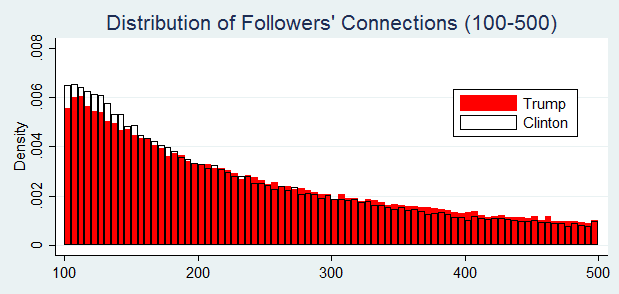}
\end{figure}

We find that individuals with only a few followers and individuals with hundreds of followers make up a larger share in the Trump camp than in the Clinton camp, while by contrast individuals with a few dozen to two hundred followers have a larger presence among the Clintonists.

\subsection{Gender}
In this subsection, we analyze the gender composition of Trump's and Clinton's followers. As reported in Section 3, we train a CNN model using the MORPH database and use the trained model to classify Twitter profile images.

A number of studies have demonstrated the ``gender affinity effect'' in American elections \cite{women4women,runAsWomen}. In the first Democratic public debate in October, 2015, Hillary Clinton also emphasized her identity as a woman: ``Being the first woman president would be quite a change from the presidents we've had, including President Obama.''\footnote{http://www.huffingtonpost.com/entry/hillary-clinton-first-woman-president\_561dbf71e4b028dd7ea5af6c.} Clinton is certainly enjoying the support of her fellow female politicians. Out of 14 female Democratic senators, 13 have endorsed Clinton's presidential campaign.\footnote{http://www.cnn.com/2015/11/30/politics/hillary-clinton-elizabeth-warren-fundraiser.} But there is also strong evidence that shows Clinton's support among average Democratic women has fallen sharply.\footnote{https://www.washingtonpost.com/politics/poll-sharp-erosion-in-clinton-support-among-democratic-women/2015/09/14/6406e2a0-58c3-11e5-b8c9-944725fcd3b9\_story.html.} This makes our investigation in the Twitter sphere particularly interesting.\\

\begin{figure}[]
\caption{Gender Composition}
\label{gender}
\includegraphics[height=4cm]{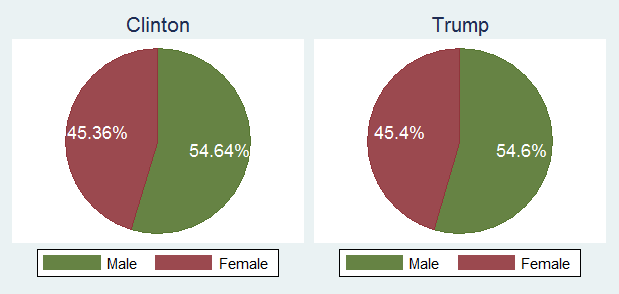}
\end{figure}

\begin{table}[H]
\centering
\caption{Score Test on Gender Composition}
\label{stgender}
\setlength{\tabcolsep}{16pt}
\begin{tabular}{lll}
\hline\hline
\multirow{2}{*}{Null Hopythesis} & \multicolumn{2}{c}{Female} \\
\cline{2-3}
                                 & z statistic     & \textit{p} value      \\\hline
p$_{c}$=p$_{t}$                            & -0.11           & 0.46    \\\hline
\end{tabular}
\end{table}

We report the findings in Figure \ref{gender} and the score test statistic in Table \ref{stgender}, where p$_c$ denotes probability among the Clintonists and p$_t$ denotes the probability among the Trumpists.\footnote{The formula for the score test statistic is: $z=\frac{\hat{p}_1-\hat{p}_2}{\sqrt{\hat{p}(1-\hat{p})(1/n_1+1/n_2)}}$, where $\hat{p}_1=\frac{x}{n_1},\:\hat{p}_2=\frac{y}{n_2},\:p=\frac{x+y}{n_1+n2}.$ With large $n_1$ and $n_2$, z is approximately standard normal.} The main reading is that in the Twitter sphere we have not detected any gender affinity effect. The percentage of female followers for Clinton is not larger than that for Donald Trump. This also means that Trump has strong female support and that apparently Trump's feud with Megyn Kelly has not alienated female voters.

\subsection{Race}
In this subsection, we analyze the racial composition of Trump's and Clinton's followers. Research of 2008 exit poll data has shown that Barack Obama won the presidential election because of his race rather than despite his race. Ethnic minorities, such as African Americans and Hispanics, tend to vote for the Democrats (Stewart and Ansolabehere, 2009). Here our goal is to explore this phenomenon in the Twitter sphere.

\begin{figure}[h!]
\caption{Racial Composition}\label{race}
\includegraphics[height=4cm]{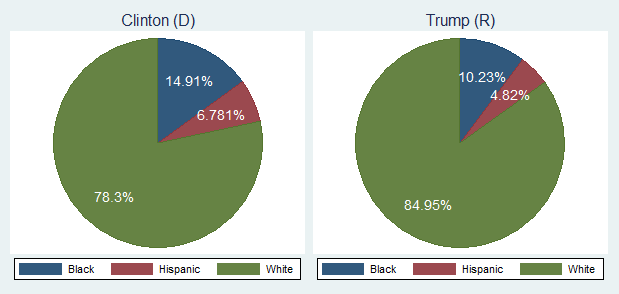}
\end{figure}

In Figure \ref{race}, we report the calculated racial composition of the Clintonists and the Trumpists. Our results show that a Clintonist is more likely to be an African American or Hispanic than a Trumpist is. A Trumpist has a higher probability to be white than a Clintonist. This pattern in Twitter sphere is consistent with historical voting patterns.

To test the significance of the differences in racial composition, we calculate score test statistic for each race and report them in Table \ref{race}.\footnote{In the test of each race, we treat it as binary.} We find these differences in followers' racial composition statistically significant.
\begin{table}[]
\centering
\caption{Score Test on Racial Composition}
\label{binomial}
\setlength{\tabcolsep}{5pt}

\begin{tabular}{lllllll}
\hline\hline
\multirow{2}{*}{Null} & \multicolumn{2}{c}{Black} & \multicolumn{2}{c}{Hispanic} & \multicolumn{2}{c}{White} \\\cline{2-7}
                                 & z statistic   & \textit{p}    & z statistic     & \textit{p}       & z statistic    & \textit{p}     \\\hline
p$_{c}$=p$_{t}$                             & 17.1        & 0         & 10.1          & 0          & -20.7       & 0       \\\hline 
\end{tabular}
\end{table}

\subsection{Age}
In this subsection, we study the age distribution of the candidates' followers. A stereotypical idea is that the Republican supporters tend to be old white people. Also for this reason, the Republican party is considered to be in a demographic crisis as its supporters are gradually passing away.

\begin{figure}[h!]
\caption{Age Distribution}
\label{age}
\includegraphics[height=4cm]{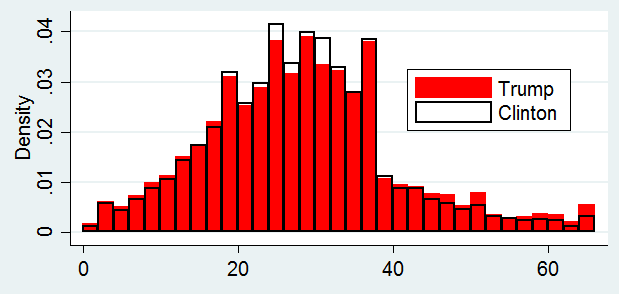}
\end{figure}

We explore this demographic question by estimating the ages of the candidates' followers. We report our results in Figure \ref{age}. Consistent with real world voting, we find that older people aged above 40 make up a larger share in the Trump camp than in the Clinton trump. We also find that followers of a very young age (1-17) also make up a larger share in the Trump camp. But notice that these very young people are not eligible for voting yet. For followers aged between 18 and 40, they occupy a larger presence among the Clintonists than among the Trumpists.

Using score test we are able to show that the Clintonists in the Twitter sphere are statistically more likely to be in the 18-40 age group than the Trumpists and less likely to be in either the 1-17 age group or the 41-66 age group (Table \ref{age_dis}).
\begin{table}[H]
\centering
\caption{Score Test of Age Composition}
\label{age_dis}
\setlength{\tabcolsep}{3.5pt}

\begin{tabular}{lllllll}
\hline\hline
\multirow{2}{*}{Null} & \multicolumn{2}{c}{1-17} & \multicolumn{2}{c}{18-40} & \multicolumn{2}{c}{41-66} \\\cline{2-7}
                                 & z statistic   & \textit{p}    & z statistic     & \textit{p}       & z statistic    & \textit{p}     \\\hline
p$_{c}$=p$_{t}$                             & -3.1        & 0.002         & 8.9          & 0          & -9.1       & 0       \\\hline 
\end{tabular}
\end{table}

\section{Conclusions} 
In this study, we have analyzed four dimensions of follower demographics: social influence, gender, race and age. With respect to social influence, we find that Trump's followers are more likely to find themselves in the tails of the distribution, while Clinton followers are more likely to appear in the middle. The same pattern also holds in the age dimension.  Our study finds no gender affinity effect for Clinton in the Twitter sphere, but we do find that the Clintonists are more racially diverse. 


\section{Acknowledgment}
We gratefully acknowledge support from the University of Rochester, New York State through the Goergen Institute for Data Science, and our corporate sponsors Xerox and Yahoo.

\bibliographystyle{aaai}
\bibliography{yu}
\end{document}